\documentclass[12pt]{iopart}
\usepackage{graphicx,bm}
\usepackage{amssymb,color}
\usepackage{iopams}
\usepackage{txfonts}
\usepackage{array}
\usepackage{graphicx,graphics,rotating}	% Include figure files
\usepackage{dcolumn}		% Align table columns on decimal point
\usepackage{bm}				% bold math

\newcommand{\be}{\begin{equation}}
\newcommand{\ee}{\end{equation}}
\newcommand{\ba}{\begin{eqnarray}}
\newcommand{\ea}{\end{eqnarray}}
\newcommand{\bra}[1]{\ensuremath{\langle{#1}|}}
\newcommand{\ket}[1]{\ensuremath{|{#1}\rangle}}

\newcommand{\matel}[3]{\ensuremath{\bra{#1}\hat{#2}\ket{#3}}}
\newcommand{\dket}[1]{\|{#1}\rangle\negthinspace \rangle}       %operator ket 		%
\newcommand{\dbra}[1]{\langle\negthinspace \langle{#1}\|}        %!operator bra
\newcommand{\dbraket}[2]
{\ensuremath{\langle\negthinspace\langle{#1}\|{#2}\rangle\negthinspace \rangle}}  %operator braket
\newcommand{\dmatel}[3]{\ensuremath{\dbra{#1}~{#2}~\dket{#3}}}   %superoperator matrix element
\newcommand{\vectr}[2]{\left(\begin{array}{lcl} {#1} \\ {#2} \end{array} \right)}
\newcommand{\tvectr}[2]{\left(\begin{array}{lclcl}{#1} & {#2} \end{array} \right)}
\newcommand{\matr}[4]{\left(\begin{array}{lclcl}{#1}&{#2}\\{#3}&{#4} \end{array} \right)}
\newcommand{\supp}[1]{\mathbf{#1}}
\newcommand{\sym}[2]{\ensuremath{<{#1},{#2}>}}    %symplectic form
\newcommand{\opS}[1]{\ensuremath{ S_{ \boldsymbol {#1}}}}  %generic basis element
\newcommand{\optr}[1]{\ensuremath{ T_{ {#1}}}}   %translation operator
\newcommand{\opref}[1]{\ensuremath{ R_{ {#1}}}}      %reflection operator
\newcommand{\opSh}[1]{\ensuremath{\hat S_{ \boldsymbol {#1}}}}  %generic basis element hat
\newcommand{\optrh}[1]{\ensuremath{ \hat T_{\boldsymbol {#1}}}}   %translation operator hat
\newcommand{\oprefh}[1]{\ensuremath{\hat R_{\mathbf {#1}}}}      %reflection operator  hat
\def\bfalpha{{\boldsymbol \alpha}}
\def\balpha{{\boldsymbol \alpha}}
\def\bfbeta{{\boldsymbol \beta}}
\def\bbeta{{\boldsymbol \beta}}

\newcommand{\x}{{\mathbf x}}  %from alfredo
\newcommand{\y}{{\mathbf y}}  %from alfredo
\newcommand{\bfa}{{\mathbf a}}

\newcommand{\opR}{{\hat{R}}}
\newcommand{\opT}{{\hat{T}}}
\newcommand{\oprho}{{\hat{\rho}}}
\def\mbfx{{\mathbf x}}
\def\mbfy{{\mathbf y}}

\newcommand{\half}{\frac{1}{2}}

\begin{document}

%\preprint{APS/123-QED}

\title{ Representation of superoperators in double phase space}

\author{Marcos Saraceno$^1$ and Alfredo M. Ozorio de Almeida$^2$}
 \address{$^1$
 Departamento de F\'isica Te\'orica, GIyA, Comisi\'on Nacional de Energ\'\i a At\'omica, Av. Libertador 8250, C1429BNP Buenos Aires, Argentina
}%
\address{$^2$Centro Brasileiro de Pesquisas Fisicas, 
Rua Xavier Sigaud 150, 22290-180, 
Rio de Janeiro, R.J., Brazil.}


\begin{abstract}
Operators in quantum mechanics - either observables, density or evolution operators, unitary or not - can be represented by c-numbers in operator bases. The position and momentum bases are in one to one correspondence with lagrangian planes in double phase space, but this is also true for the well known Wigner-Weyl correspondence based on translation and reflection operators. These phase space methods are here extended to the representation of superoperators. We show 
that the Choi-Jamiolkowsky isomorphism between the dynamical matrix and the linear action of the superoperator 
constitutes a "double" Wigner or chord transform when represented in double phase space. As a byproduct several previously unknown 
integral relationships between products of Wigner and chord distributions for pure states are derived. 
\end{abstract}

\maketitle
%%Introduction

%%%%%%%%%%%%%%%%%%%%%%%%%%%%%%%%%%%%%%%%%%%%%%%%%
\section{Introduction}

The action of a unitary operator on states in quantum Hilbert space
corresponds semiclassically to a classical transformation, $\x_- \mapsto \x_+$, 
in the corresponding classical phase space ${\mathbb R}^{2N}$, 
with points $\x=(q_1,..., q_N, p_1,..., p_N)$. Alternatively,
one may place the points $\x_+$ and $\x_-$  each in its own phase space
so that the graph of a canonical transformation becomes a $2N-$ dimensional ($2N-$D) surface
within the {\it double phase space}, built from the product space ${\mathbb R}^{2N}\times{\mathbb R}^{2N}$, 
with points $X=(\x_+,\x_-)$ \cite{Marsden}.

In strict analogy to the semiclassical correspondence of an (integrable) quantum state
with a classical $N$-D surface in ${\mathbb{R}}^{2N}$ \cite{Gutzwiller, livro},
evolution of the operator corresponds classically to movement of the surface in its phase space.
Thus, a coordinate transformation in phase space, $\x_- \mapsto \x_+$, implies a mapping 
$X \mapsto X'$, in double phase space 
(for instance, a {\it normal form transformation} \cite{Arnold, livro})
that may be seen to propagate the classical surface for a given canonical transformation 
in double phase space, corresponding to the action of a {\it superoperator} 
as it evolves ordinary operators in quantum mechanics.

One should keep in mind some essential differences between the classical 
and the quantum scenarios. Above all there is the uncertainty principle:
while vectors in Hilbert space correspond semiclassically to (complex) functions on the lagrangian manifolds on single phase space - say the position {\sl or} the momentum basis wave functions- operators are represented by their matrix elements from a pair of such manifolds. When viewed in double phase space  these pairs are themselves $2N$-D lagrangian surfaces in $4N$-D double phase space.

However, some special coordinate transformations are allowed
that take these obvious lagrangian coordinate planes of double phase space, 
that is, initial and final positions or momenta,
to new coordinate planes that cannot be so decomposed. This is just the case
of phase space labels underlying the Weyl representation (i.e. the Wigner function
in the case of the density operator) and its Fourier transform (FT).  
Thus, a unitary operator $\hat U$  corresponding to a given surface in double phase space
is represented by $U(Q)=\langle q_+|U|q_-\rangle$, $U(P)=\langle p_+|U|p_-\rangle$,
or in the Weyl representation as $U(q,p)$, depending on rotations in double phase space
that correspond to different choices of FT's in quantum mechanics.

But the analogy of an operator represented by a lagrangian surface in double phase space to the state represented by a lagrangian surface in simple phase space
 can now be pushed a step further: What about the 'double Weyl transform' of
a superoperator, represented as a function in double phase space?
The answer invokes the adaptation of the {\it Choi-Jamiolkowsky isomorphism} \cite{jamio}, well known in the theory of quantum information, 
to the operator basis underlying the Weyl representation and its Fourier transform.
The purpose of this paper is to undertake this adaptation, showing that the Choi or dynamical matrix of a superoperator is the 'double Weyl transform' of its matrix elements in the Weyl basis. We achieve this by defining reflection and translation superoperators, labeled by points in double phase space, in strict analogy to the corresponding definitions in single phase space. Moreover we show that they take a simple monomial form in the Choi conjugate basis.

Another surprise is that the Choi-conjugate pairs of operators representing
 pure states are identical. This leads to unexpected Fourier
identities involving integrals of products of ordinary Wigner functions. In some cases this in turn leads to new identities involving the special functions of analysis. 
 
The paper is organized as follows: in section 2 we review the concept of double phase space \cite{Marsden, Amiet, Littlejohn, LNP} and how it is related to the labeling of conjugate bases of operators in terms of lagrangian surfaces in double phase space.  

In section 3 we extend the methods to the representation of superoperators. 
The concept of Choi-conjugate bases is defined and applied to the phase space bases 
that are determined by reflections and translations. In this context Choi-conjugation 
takes on the character of a ``double'' Wigner -Weyl transformation in double phase space. 
The identification of such double Wigner functions with the Wigner-Weyl transformation 
from alternative matrix representations of a superoperator is examined in section 4.
 
Superoperators for various kinds of quantum evolution are then discussed in section 5. 
Finally, in section 6 the pullback of the double phase space results for single phase space
Wigner functions generates several new identities for pure states.

%%%%%%%%%%%%%%%%%%%%%%%%%%%%%%%%%%%%%%%%%%%%%%
\section{Double phase space and operator representations}
%%%%%%%%%%%%%%%%%%%%%%%%%%%%%%%%%%%%%%%%%%%%
The concept of double phase space underlies in classical mechanics the general theory of generating functions  of canonical transformations. It provides an elegant way to visualize a canonical transformation in a doubled phase space as the gradient of a generating function \cite{Marsden, Amiet, Littlejohn}. In quantum mechanics it provides a flexible mechanism to represent unitary  propagators in the semiclassical limit \cite{Miller} as functions on these manifolds and underlies the presentation of quantum mechanics in terms of phase space path integrals \cite{Feynmann,Schulman}. 

 A canonical transformation  \footnote{ For simplicity from now on we restrict to the 1-D case $\mbfx=(q,p)$} $\mbfx_-\mapsto\mbfx_+$ can be specified implicitly by a generating function $S(q_+, q_-)$ whose differential is 
\be
\rmd S(q_+,q_-)=p_+\rmd q_+ -p_-\rmd q_-
\ee
 With a simple redefinition of coordinates
\be
  Q_1=(q_+,q_-)  ~~~~~~~~~P_1=(p_+,-p_-)
\label{Q1}  
\ee
we rewrite  $\rmd S=P_1\cdot\rmd Q_1 $ and the transformation is defined implicitly as $P_1(Q_1)=\partial S(Q_1)/\partial Q_1$. The new coordinates $Q_1,P_1$ can now be interpreted as canonical coordinates in a phase space with doubled dimensions. In the elementary theory other canonical pairs are well known and are obtained by various Legendre transforms. In the standard notation of \cite{Goldstein, Synge, Jose}  they are:
\ba
\rmd S =\rmd F_1 &=&p_+\rmd q_+ - p_-\rmd q_- \equiv P_1\cdot\rmd Q_1  ~~~~~~~Q_1= (q_+,q_-),    ~~~~P_1=(p_+,-p_-)\nonumber\\
\rmd F_2 &=& q_+\rmd p_+ + p_-\rmd q_-\equiv P_2\cdot\rmd Q_2 ~~~~~~~Q_2= (p_+,q_-),    ~~~~P_2=(q_+,p_-)\nonumber\\
\rmd F_3 &=&- q_-\rmd p_- - p_+\rmd q_+ \equiv -Q_2\cdot\rmd P_2 \nonumber\\
\rmd F_4 &=&- q_-\rmd p_- + q_+\rmd p_+ \equiv -Q_1\cdot\rmd P_1 
\label{classgenfun}
\ea
The general theory developed by  \cite{Marsden, Amiet, Littlejohn} interprets these alternative pairs $Q_j, P_j$ as the coordinates for hyperplanes lying in a  phase space with doubled dimension.
 We give here a synthetic overview with the purpose of identifying the structures that will reappear in quantum mechanics in the representation of operators and superoperators.
Starting from the elementary statement of area preservation in a canonical transformation 
\be
\oint p_+\rmd q_+=\oint p_-\rmd q_-
\ee
rewritten as
\be
\half\oint \mbfx_+ \cdot J \rmd \mbfx_+ -\mbfx_-\cdot J \rmd  \mbfx_- =0 ,
\label{areapreserv}
\ee
where as usual $\mbfx_\pm\equiv (q,p)_{\pm}$ are points in single phase space and 
$J=\matr{0}{-1}{1}{0}$ is the standard symplectic form in $2$D. Placing ourselves in the $4$D direct product space spanned by $(\mbfx_+ , \mbfx_-)$ we rewrite \eref{areapreserv} as
\be
\half\oint \tvectr{\mbfx_+}{\mbfx_-}\matr{J}{0}{0}{-J}\vectr{\rmd \mbfx_+}{\rmd \mbfx_-}=0 .
\ee
Thus the canonical transformation $ \mbfx_-\mapsto\mbfx_+$ is defined by a $2$D lagrangian surface in the $4$D direct product phase space. On this surface
\be
\rmd A= \half\tvectr{\mbfx_+}{\mbfx_-}\matr{J}{0}{0}{-J}\vectr{\rmd \mbfx_+}{\rmd \mbfx_-}
\ee
is an exact differential. However this surface is lagrangian with respect to the non standard symplectic form $\matr{J}{0}{0}{-J}$. To introduce canonical coordinates with respect to the standard canonical form $\matr{0}{-1}{1}{0}$, we consider a linear transformation 
\be
\vectr{\mbfx_+}{\mbfx_-}= {\cal U}\vectr{Q}{P}\equiv\matr{U_{00}}{U_{01}}{U_{10}}{U_{11}}\vectr{Q}{P}
\label{canonical}
\ee
such that
\be
\rmd A=\half ( \mbfx_+ \cdot J \rmd \mbfx_+ -\mbfx_-\cdot J \rmd  \mbfx_-) =
\half (P \cdot\rmd Q -Q\cdot\rmd P).
\label{xytoQP}
\ee
The matrix $\mathcal U$ should satisfy
\be
{\mathcal U}^t \matr{J}{0}{0}{-J}{\mathcal U}=\matr{0}{-1}{1}{0}.
\label{condition}
\ee
Various choices of the matrix ${\cal U}_i$ lead to different pairs $Q_i,P_i$, and all such pairs are related by canonical transformations in 4-D phase space.
Notice that the first form of $\rmd A$ in \eref{xytoQP}  
 is invariant under canonical transformations $\mbfx_\pm \to \mbfy_\pm $ in single phase space while the second form  becomes invariant under full two degrees of freedom canonical transformations $ Q,P \to Q',P'$ .
To proceed to the definition of the canonical transformation in these coordinates a Legendre transformation on $\rmd A$ leads to
\be
\rmd \bar A=\rmd (A +\half P\cdot Q)=P\cdot\rmd Q
\label{genfun}
\ee
and therefore
\be 
P(Q)=\frac{\partial\bar A(Q)}{\partial Q}.
\ee
 The canonical transformation is then specified via the parametric equations
\be 
\vectr{\mbfx_+ (Q )}{\mbfx_-(Q)}
=\matr{U_{00}}{U_{01}}{U_{10}}{U_{11}}
 \vectr{Q}{P (Q)} 
 \label{param1}
\ee 
Alternatively a different Legendre transform leads to a conjugate representation exchanging coordinates and momenta $\rmd\tilde{ A}(P)=\rmd (A-\half P\cdot Q)=-Q\cdot\rmd P$ leading to $Q=-\partial\tilde A/\partial P$ and to the parametric equations
\be 
\vectr{\mbfx_+ (P )}{\mbfx_-(P)}
=\matr{U_{00}}{U_{01}}{U_{10}}{U_{11}}
 \vectr{Q(P)}{P } 
 \label{param2}
\ee 
In this general approach a canonical transformation is characterized by a) a linear transformation $\mathcal U$ satisfying \eref{condition} and defining the canonical coordinates $Q,P$ and b) by generating functions $\bar A(Q)$  ($\tilde A(P)$) on the lagrangian surfaces $P=const. $ ($Q=const.$). The two generating functions are themselves related by  the Legendre transform $\bar A = \tilde A + P\cdot Q$. Singularities in the process of inverting the functions $\mbfx_-(Q)$ ($\mbfx_+(P)$) in \eref{param1} and \eref{param2} will lead to transversality conditions guaranteeing the existence of the transformation. This general scheme can be easily identified in \eref{classgenfun} where $ Q_1, P_1$ and $Q_2, P_2 $ are canonically conjugate pairs and $\mathcal U$ is a simple permutation.  

A different choice of canonical variables in double phase space 
is related to the well known Wigner-Weyl representation of quantum mechanics in phase space. 
It is characterized by the matrix
 \be
 {\mathcal U}=\matr{1}{\half J}{1}{-\half J},
 \ee
 leading to the transformation
 \be
 \mbfx_+=Q +\half J P ~~~~~~~~~~~ \mbfx_-=Q -\half J P
 \label{xtoQP}
 \ee
 and its inverse
 \be
 Q = \frac{\mbfx_+ +\mbfx_-}{2} ~~~~~~~~~~~~ P=-J(\mbfx_+-\mbfx_-).
 \label{QPtox}
 \ee
 The lagrangian planes of constant $Q$ represent the transformation $\x_+=2Q-\x_-$ which is a phase space reflection, while those of constant $P$ lead to the translation $\x_+=\x_- +JP$. Thus $Q,P$ are natural labels in quantum mechanics for the operators that represent translations and reflections. However it is customary to use, instead of $(Q,P)$, the single phase space coordinates
 \be
 \mbfx\equiv Q=(\mbfx_++\mbfx_-)/2~~~~~~~~~\bxi\equiv JP=\mbfx_+-\mbfx_ .
 \label{transf}
 \ee
They have a simple geometrical interpretation in single phase space as centers and chords of a pair of phase space points \cite{Report} and the Poincar\'e generating function, $S(\x)$ has many desirable 
properties \cite{Weinstein, Report}.
 
 Lagrangian coordinates in double phase space provide the natural labels of operator bases in quantum mechanics. In the simple case of the position basis $\ket{q}$ an operator is represented by its matrix elements $\matel{q_+}{O}{q_-}=\tr\hat O \ket{q_-}\bra{q_+}$ as a function on the lagrangian plane $Q_1=(q_+,q_-)$. In the conjugate momentum basis $\ket{p_+}\bra{-p_-}$ the same is true as a function on $P_1=(p_+,p_-)$.
 
To treat the general case we consider a basis of operators $\opSh{\balpha}$ labeled by a double index $\balpha\in\mathbb{R}^2$ which can be any of the lagrangian planes considered. Introducing the notation $\dbraket{A}{B}\equiv \tr \hat A^\dagger \hat B$ for the Hilbert Schmidt scalar product, we adopt a dual notation for operators: when they are considered as vectors in the Hilbert-Schmidt sense we use the double Dirac notation $\dket{ A}$, but when we consider them as two sided arrays with the standard multiplication rules they will be denoted simply as $\hat A$. The latter notation emphasizes the active aspect of $\hat A$ as operators, while the former is meant to bring out their passive role as basis elements. The bases corresponding to the lagrangian planes in \eref{classgenfun} are 
 $\dket{Q_1}=\ket{q_+}\bra{q_-}$,$~\dket{P_1}=\ket{p_+}\bra{-p_-}$,$~\dket{Q_2}=\ket{p_+}\bra{q_-}$,$~\dket{P_2}=\ket{q_+}\bra{p_-}$. Their complementarity is reflected in the property $\dbraket{P_i}{Q_i}=\exp -\rmi Q_i\cdot P_i/\hbar$.
 The common properties of these  bases are
\ba
   \dbraket{\opS{\alpha}}{\opS{\beta}}=\Lambda\delta(\bfalpha-\bfbeta)  ~~~~~~~~~~~&\mathrm{Orthogonality}\\
   \frac{1}{\Lambda}\int \rmd^2\bfalpha~\dket{\opS{\alpha}}\dbra{\opS{\alpha}}= 1    ~~~~~~~~~~~~& \mathrm{Completeness}
   \label{completeness}
\ea
One should note that, within the context of the following section, the latter relation actually represents the {\it identity superoperator}.
The factor $\Lambda$ is conventional and could be removed by a simple rescaling. 
Eq.\eref{completeness} implies that any operator $\hat O$ can be expanded in the basis $\dket{\opS{\alpha}}$, i.e.
\be
\hat O=\frac{1}{\Lambda}\int \rmd^2\bfalpha~\opSh{\alpha}\dbraket{\opS{\alpha}}{O},
\ee
with coefficients $\dbraket{\opS{\alpha}}{ O}$ that are the c-number representatives of $\hat O$ . In the simple ket, bra bases considered above they are simply its matrix elements. 
 
 Another pair of complementary operator bases is obtained when considering the unitary representation of reflections and translations in phase space. They are defined in terms of symmetrized Fourier transforms of the previous bases as follows
\ba
\oprefh{\x}\equiv\hat R_{q,p} &=&\int \rmd q' ~\ket{q+\frac{q'}{2}}\bra{q-\frac{q'}{2}}~\omega^{pq'} \label{transx}\\
\optrh{\bxi}\equiv\hat T_{\xi_q,\xi_p} &=&\int \rmd p' ~\ket{p'+\frac{\xi_p}{2}}\bra{p'-\frac{\xi_p}{2}}~\omega^{-\xi_q p'} 
\label{refx}
\ea
Here the first is in the position basis and the latter in the momentum basis, but either can be used. 
\footnote{It should be noted that the unitary reflection operator is $\oprefh{\x}/2$, but
our definition is more convenient as a basis.}
To simplify notation we have defined $\omega\equiv\exp \rmi/\hbar$. As a complementary set of operator bases they have the properties
\be
    \dbraket{\opref{\x}}{\opref{\x'}}=2\pi\hbar\delta(\x-\x'),~~~
    \dbraket{\optr{\bxi}}{\optr{\bxi'}}=2\pi\hbar\delta(\bxi-\bxi'),~~~
    \dbraket{\optr{\bxi}}{\opref{\x}}=\omega^{-\sym{\bxi}{\x}}
\ee
so that for these bases $\Lambda= 2\pi\hbar$, whereas $\Lambda=1$ for the position or momentum bases. The action of these operators and further properties are reviewed in the Appendix. 

The c-number representatives of operators in these bases are phase space functions 
that constitute the starting point for the formulation of quantum mechanics in phase space \cite{Wigner,Groenewold,Moyal,BalazsJen}. The reflection basis, 
 \be
O (\x)\equiv\dbraket{\opref{\mbfx}}{\hat O} = \int\rmd q' ~\matel{q+\frac{q'}{2}}{O}{q-\frac{q'}{2}}~\omega^{-q' p}, 
\label{center}
\ee
 leads to the Wigner- Weyl symbol of the operator while the translation basis
\be
\tilde O(\bxi)\equiv\dbraket{\optr{\bxi}}{\hat O}=\int\rmd p' ~\matel{p'+\frac{\xi_p}{2}}{O}{p'-\frac{\xi_p}{2}}~\omega^{p' \xi_q}, 
\label{chord}
\ee
leads to its characteristic function. In line with \cite{Report} we will refer to them respectively as centre and chord representations.
They are related by a symplectic Fourier transform
\be
O (\x)=\frac{1}{2\pi\hbar}\int \rmd \bxi^2~ \omega^{-\sym{\x}{\bxi}}~ \tilde O(\bxi).
\ee
 In the special case of the density operator $\oprho$, $\rho(\x)$
and $\rho(\xi)$ are the Wigner function and the characterisitc function, or chord function:
\be
\chi(\xi)=\frac{1}{2\pi\hbar}\dbraket{\optr{\xi}}{\rho}~~~~~~~{\rm and}~~~~~~~~~W(\x)=\frac{1}{2\pi\hbar}\dbraket{\opref{\x}}{\rho},
\label{defwigchord}
\ee
with the special normalization properties
\be
2\pi\hbar~\chi(0)=\tr\hat \rho=1~~~~~~~{\rm and}~~~~~~~~~\int\rmd\x^2~ W(\x)=\tr\hat \rho=1,
\label{defwig}
\ee
corresponding to classical distributions.

%%%%%%%%%%%%%%%%%%%%%%%%%%%%%%%%%%%%%%%%%%%%%%%%%%%%%%%%%
\section{Superoperator representations}

\subsection{General representations and the Choi- Jamiolkowsky isomorphism}
  A superoperator $\supp{S}$ is a general linear transformation in the space of operators. Simple examples are the unitary evolution of a density matrix $ \hat \rho\to \hat U\hat\rho\hat U^\dagger $, symmetry operations, time reversal, non-unitary evolution, etc. Once the action of $\supp{S}$ on general operators has been specified as $\supp{S}(\hat O)=\hat O'$, its matrix elements in the $\opS{\bfalpha}$ basis are 
\be
\dmatel{ \opS{\alpha}}{\supp{S}}{\opS{\beta}}=\tr [{\opSh{\alpha}}^\dagger\supp{S}(\opSh{\bfbeta})]
\label{suppmat}
\ee
Thus a general superoperator can be expanded as
\be
\supp{ S}=\frac{1}{\Lambda^2} \int \rmd^2\bfalpha\rmd^2\bfbeta ~
\dmatel{ \opS{\bfalpha}}{\supp{S}}{\opS{\bfbeta}} \dket{\opS{\alpha}}\dbra{\opS{\beta}}.
\label{supp}
\ee
The superoperators $\dket{\opS{\alpha}}\dbra{\opS{\beta}}$ constitute then a complete orthonormal basis labeled by a double index $\bfalpha,\bfbeta \in\mathbb{R}^4$. 
Their action on an operator ${\hat O}$ follows the standard bra and ket rules so that 
$\dket{\opS{\alpha}}\dbraket{\opS{\beta}}{O}\equiv \opSh{\alpha}\tr\opSh{\beta}^\dagger \hat O$.
The unitary relations between different choices of bases also allow us to define the trace of a superoperator uniquely as
\be
\Tr~ \supp{ S} \equiv \frac{1}{\Lambda} \int \rmd^2\bfalpha ~
\dmatel{ \opS{\bfalpha}}{\supp{S}}{\opS{\balpha}}.
\ee

Nonetheless, it is convenient to consider a different action related to the pair of operators $\opSh{\alpha}, {\opSh{\beta}}^\dagger$  
\be
\opSh{\alpha}\bullet {\opSh{\beta}}^\dagger ~ (\hat O)\equiv
\opSh{\alpha}~\hat O~ \opSh{\beta}^{\dagger}.
\ee
Indeed, $ \opSh{\alpha}\bullet {\opSh{\beta}}^\dagger $ is a different orthonormal superoperator basis related to the former by a unitary transformation in $ \mathbb{R}^4$. The transformation is given by
\be
\opSh{x}\bullet\opSh{y}^\dagger=\frac{1}{\Lambda^2} \int \rmd^2\bfalpha~\rmd^2\bfbeta~ \tr(\opSh{x}\opSh{\beta}\opSh{y}^\dagger\opSh{\alpha}^\dagger)~\dket{\opS{\alpha}}\dbra{\opS{\beta}},
\label{rel}
\ee 
with the inverse:
\be
\dket{\opS{\alpha}}\dbra{\opS{\beta}}=\frac{1}{\Lambda^2} \int \rmd^2\mbfx~\rmd^2\mbfy~\tr(\opSh{x}^\dagger\opSh{\alpha}\opSh{y}\opSh{\beta}^\dagger)~\opSh{x}\bullet\opSh{y}^\dagger.
\label{relinv}
\ee
We prove for example \eref{rel} by computing with \eref{suppmat} the matrix element $\dmatel{{\opS{\alpha}}}{\opSh{x}\bullet\opSh{y}^\dagger}{\opS{\beta}}=\tr(\opSh{x}\opSh{\beta}\opSh{y}^\dagger\opSh{\alpha}^\dagger)$. In the following we will refer to the pair of superoperators $\dket{A}\dbra{B}$ and $\hat A\bullet\hat B^\dagger$  as Choi-conjugates and extend this name to the two bases. 

$\supp{S}$ can also be expanded in the new basis as
\be
\supp{S}=\frac{1}{\Lambda^2}\int \rmd^2\mbfx\rmd^2\mbfy \, C_\supp{S}(\mbfx,\mbfy)\,\opSh{x}\bullet\opSh{y}^\dagger,
\label{suppchoi}
\ee
so that the coefficients $C_\supp{S}(\mbfx,\mbfy)$ constitute an alternative representation of $\supp{S}$, related to its matrix elements by \eref{relinv}
\be
\dmatel{ \opS{\bfalpha}}{\supp{S}}{\opS{\bfbeta}}=\frac{1}{\Lambda^2}\int \rmd^2\mbfx\rmd^2\mbfy \, C_\supp{S}(\mbfx,\mbfy)\,\tr(\opSh{\alpha}^\dagger\opSh{x}\opSh{\beta}\opSh{y}^\dagger). 
\label{suppmatel}
\ee

The nature of the transformation between the two representations of $\supp{S}$ is specific of a given basis and depends on the trace of four basis elements. The simplest transformation occurs when a transition basis $\hat E_{ij}=\ket{i}\bra{j}$ of orthonormal states is used. In that case (setting $\Lambda=1$) the quadruple trace results in a product of four delta functions which reduce the relationship to a partial transposition of indices
\be
\dmatel{E_{ij}}{\supp{S}}{E_{kl}}=C_{\supp{S}}(ik,jl).
\ee
In this case one obtains two different linear transformations related to 
$\supp{S}$. They are well known 
 in the quantum open systems literature. $ C_{\supp{S}}(ik,jl)$  is known as the Choi matrix \cite{Choi}, dynamical matrix \cite{BengZycs,Sudarshan}, chi matrix \cite{N-C} and the relationship between them as the Choi- Jamiolkowski isomorphism. It is also of importance in the literature on matrix product states for spin chains \cite{Rommer}. When $\supp{S}$ represents a quantum operation on density matrices the physical requirement of complete positivity requires that $C_{\supp{S}}(ik,jl) $ be hermitian and positive \cite{Kraus,Choi}. In that case $ C_{\supp{S}}$ maps the space of positive density matrices into itself. Although related, the spectral properties of the two transformations are very different. Thus if $C_{\supp{S}}(ik,jl) $ is hermitian it can be brought to diagonal form leading to the Kraus representation \cite{Kraus}. On the other hand $\dmatel{E_{ij}}{\supp{S}}{E_{kl}}$ as a linear transformation may not even be normal, leading in general to a complex spectral decomposition.

%%%%%%%%%%%%%%%%%%%%%%%%%%%%%%%%%%%%%%%%%

\subsection{The Choi-Jamiolkowsky relation in phase space}

Consider first the reflection basis $\oprefh{\x}$. A general superoperator $\supp{S}$ in this basis 
has the matrix elements $\dmatel{\opref{\x_{+}}}{\supp{S}}{\opref{\x_{-}}}$. 
The labels, $\x_{\pm}$ in $\mathbb{R}^2$, span a full set of reflection centres, 
each covering a full phase space. On the other hand a rotation in double phase space allows us to consider 
$\dket{\opref{\mbfx_{+}}}\dbra{\opref{\mbfx_{-}}}$ as an {\it alternative position basis} for superoperators, just as valid as $\ket{q_{+}}\bra{q_{-}}$ in the operator case. Hence, one defines a {\it double Weyl transform} in analogy to \eref{center} as a symmetrized Fourier transform \footnote { The denominator $2\pi\hbar=\Lambda$
is usually absent because $\Lambda=1$ for the position representation.}
\be
\supp{S}(\x,\bxi) \equiv \frac{1}{2\pi\hbar}\int \rmd^2 \x_1
\dmatel{\opref{\x +\half \x_1}}{\supp{S}}{\opref{\x -\half \x_1}} \omega^{\sym{\x_1}{\bxi}},
\label{dweyl1}
\ee
where we use the phase space coordinates $(\x,\bxi)$ which change the Fourier phase from $-Q\cdot P$ to $\x_1\cdot J\bxi\equiv \sym{\x_1}{\bxi}$. 
This leads to a very natural definition of a reflection superoperator in this basis as
\be
\supp{R}_{\x,\bxi} \equiv
\frac{1}{2\pi\hbar}\int \rmd^2 \mbfx_1 ~ \dket{\opref{\mbfx+\half\mbfx_1}}\dbra{\opref{\mbfx-\half\mbfx_1}}
~\omega^{- \sym{\mbfx_1}{\bxi}}.
    \label{suprefQP}
\ee 
The Weyl transform of $\supp{S}$ can now be expressed as a superoperator trace,
\be
\supp{S}(\x,\bxi)=\Tr \supp{S}\supp{R}_{\x,\bxi},
\ee
in complete analogy to \eref{center}. The ``super'' reflection properties of $\supp{R}_{\x,\bxi}$ are contained in the easily derived actions on translations and reflections,
\ba
\supp{R}_{\x,\bxi}\dket{\opref{\x_0}}&=&4\dket{\opref{2\x-\x_0}}\omega^{2\sym{\x-\x_0}{\bxi}}\nonumber\\
\supp{R}_{\x,\bxi}\dket{\optr{\bxi_0}}&=& 4\dket{\optr{2\bxi-\bxi_0}}\omega^{-2\sym{\bxi-\bxi_0}{\x}},
\label{actR}
\ea
in obvious similarity to \eref{actrefl}.

Just as any other superoperator, one can now expand $\supp{R}_{\x,\bxi}$ in the Choi conjugate basis, $\oprefh{\x_1}\bullet\oprefh{\x_2}$, using the general relationship in \eref{relinv}:
\be
 \dket{\opref{\x+\half\x_1}}\dbra{\opref{\x-\half\x_1}}=\frac{1}{(2\pi\hbar)^2} \int \rmd^2\bfalpha~\rmd^2\bfbeta~
\oprefh{\balpha}\bullet\oprefh{\bbeta}~\tr(\opref{\balpha}\opref{\x+\half\x_1}\opref{\bbeta}\opref{\x-\half\x_1}).
\ee
The quadruple trace computed in \eref{traces} reduces this to
\ba
\dket{\opref{\x+\half\x_1}}\dbra{\opref{\x-\half\x_1}}&=&\frac{1}{2\pi\hbar} \int \rmd^2\bfalpha~\rmd^2\bfbeta~\oprefh{\balpha}\bullet\oprefh{\bbeta}~
\delta(\x-\frac{\balpha+\bbeta}{2})\omega^{\sym{\x_1}{\balpha-\bbeta}}\nonumber\\
&=& \frac{1}{2\pi\hbar} \int \rmd^2\bxi~~\oprefh{\x+\half\bxi}\bullet\oprefh{\x-\half\bxi}~\omega^{\sym{\x_1}{\bxi}}
\ea
and its inverse
\be
\oprefh{\x+\bxi/2}\bullet\oprefh{\x-\bxi/2}
=\frac{1}{2\pi\hbar}\int\rmd^2 \x_1 ~\dket{\opref{\x+\half\x_1}}\dbra{\opref{\x-\half\x_1}}\omega^{-\sym{\x_1}{\bxi}}.
\label{choiref}
\ee
Comparison with \eref{suprefQP} then shows that $ \supp{R}_{\x,\bxi}$ has a very simple (monomial) form in the Choi-conjugate basis:
\be
\supp{R}_{\x,\bxi}=\oprefh{\x+\bxi/2}\bullet\oprefh{\x-\bxi/2} = \oprefh{\x_+}\bullet\oprefh{\x_{-}}.
\label{suprefx}
\ee

Very similar results can be obtained in the translation basis. 
A translation superoperator is defined as
\be
\supp{T}_{\x,\bxi}\equiv\frac{1}{2\pi\hbar}\int \rmd^2 \bxi_1~
\dket{\optr{\bxi_1 +\half\bxi}}\dbra{\optr{\bxi_1 -\half \bxi}}~~
\omega^{-\sym{\bxi_1}{ \x}},
\ee
with the translation properties
\ba
\supp{T}_{\x,\bxi}\dket{\optr{\bxi_0}}=
\dket{\optr{\bxi+\bxi_0}}~~\omega^{\sym{\x}{\bxi_0+\half\bxi}}\nonumber\\
\supp{T}_{\x,\bxi}\dket{\opref{\x_0}}=\dket{\opref{\x_0+\x}}~~\omega^{\sym{\bxi}{\x_0+\half\x}}.
\label{actT}
\ea
It defines the double chord representation of $\supp{S}$ as
\be
\widetilde{\supp{S}}(\x,\bxi) \equiv \frac{1}{2\pi\hbar}\int \rmd^2 \bxi_1
\dmatel{\optr{\bxi_1 + \half \bxi}}{\supp{S}}{\optr{\bxi_1 -\half \bxi}}~~\omega^{-\sym{\bxi_1}{\x}}
=\Tr \supp{S}\supp{T}_{\x,\bxi}^\dagger.
\ee
An almost identical derivation as in \eref{choiref} yields the relationship between the Choi-conjugate translation bases:
\ba
\dket{\optr{\bxi_1+\half\bxi}}\dbra{\optr{\bxi_1-\half\bxi}}&=&\frac{1}{2\pi\hbar}\int\rmd^2\x~
\optrh{\x+\half\bxi}\bullet \optrh{\x-\half\bxi}^\dagger~~\omega^{\sym{\bxi_1}{\x}}
\nonumber\\
\optrh{\x+\half\bxi}\bullet \optrh{\x-\half\bxi}^\dagger&=&\frac{1}{2\pi\hbar}\int\rmd^2\bxi_1~
\dket{\optr{\bxi_1+\half\bxi}}\dbra{\optr{\bxi_1-\half\bxi}}~~\omega^{-\sym{\bxi_1}{\x}},
\label{choitr}
\ea
which again lead to a monomial representation of the translation superoperator:
\be
\supp{T}_{\x,\bxi}=\optrh{\x+\half\bxi}\bullet\optrh{\x-\half\bxi}^\dagger
=\optrh{\x_{+}}\bullet\optrh{\x_{-}}^\dagger.
\label{suptransx}
\ee

Equations \eref{suprefx},\eref{suptransx} are our main general results: in double phase space the reflection and translation superoperators can be defined as usual in terms of symmetrized Fourier transforms of matrix elements, but they also aquire a simple monomial form in their respective Choi-conjugate basis. The interplay between their action in double phase space with the single phase space factor translations and reflections in $\oprefh{\x_{+}}\bullet\oprefh{\x_{-}}$ and
$\optrh{\x_{+}}\bullet\optrh{\x_{-}}^\dagger$  will be further discussed in section 5.
 
We now use these identities to derive the Choi-Jamiolkowsky relationship between 
this pair of conjugate bases that represent $\supp{S}$. 
Expanding a general superoperator as in \eref{suppchoi}
\be
\fl \supp{S}=\int \frac{\rmd^2\x_+ \rmd^2\x_-}{(2\pi\hbar)^2} \, {C}_\supp{S}(\x_+,\x_-)\,\oprefh{\x_+}\bullet\oprefh{\x_-}
=\int \frac{\rmd^2\x_+ \rmd^2\x_-}{(2\pi\hbar)^2} \, \widetilde{C}_\supp{S}(\x_+,\x_-)\,\optrh{\x_+}\bullet\optrh{\x_-}^\dagger,
\ee
defines $ {C}_\supp{S}(\x_+,\x_-)$ and $\widetilde{C}_\supp{S}(\x_+,\x_-)$ as the Choi or dynamical matrices of $\supp{S}$ in the reflection and translation basis. 
Changing the double phase space variables from $(\mbfx_+,\mbfx_-) \to (\x,\bxi)$ 
as in \eref{transf} (with unit jacobian) we rewrite it as
\be
\fl \supp{S}=\int\frac{ \rmd^2 \x \rmd^2 \bxi}{(2\pi\hbar)^2} \, C_\supp{S}(\x+\frac{\bxi}{2} ,\x-\frac{\bxi}{2})\,\supp{R}_{\x,\bxi}
=\int\frac{\rmd^2 \x \rmd^2 \bxi}{(2\pi\hbar)^2} \,\widetilde{ C}_\supp{S}(\x+\frac{\bxi}{2} ,\x-\frac{\bxi}{2})\,\supp{T}_{\x,\bxi},
\label{choitrans}
\ee
that is, the superoperator is expanded as a superposition of reflections (translations), such that 
the coefficients are identified with the center (chord) representations in double phase space. 
Thus, we have obtained
\ba
\supp{S}(\x,\bxi)&=&\Tr(\supp{S}\supp{R}_{\x,\bxi})={ C}_\supp{S}(\x_+,\x_-) \label{centerpm}\\
\widetilde{\supp{S}}(\x,\bxi)&=&\Tr(\supp{S}\supp{T}^\dagger_{\x,\bxi}) =\widetilde{ C}_\supp{S}(\x_+,\x_-),
\label{chordpm}
\ea
which is the desired Choi-Jamiolkowsky relation in these bases: the Choi matrix appears as a rotated double Weyl or chord transform of the superoperator. The property can be inverted to retrieve the matrix elements from an integral over the Choi matrix
\ba
\dmatel{\opref{\x+\half\x_1}}{\supp{S}}{\opref{\x-\half\x_1}}&=& \frac{1}{2\pi\hbar}\int \rmd^2 \bxi~C_\supp{S}(\x+\half\bxi,\x-\half\bxi)\omega^{-\sym{\x_1}{\bxi}}\nonumber\\
\dmatel{\optr{\bxi_1+\half\bxi}}{\supp{S}}{\optr{\bxi_1-\half\bxi}}&=& \frac{1}{2\pi\hbar}\int \rmd^2 \x~\widetilde C_\supp{S}(\x+\half\bxi,\x-\half\bxi)\omega^{-\sym{\bxi_1}{\x}}.
\label{invref}
\ea

It is important to note that these relationships follow directly from \eref{suppmatel}, 
but the direct derivation would then miss the interpretation of the Choi matrix 
as a phase space distribution in double phase space. Furthermore, the complete analogy
between the Weyl transform in double phase space to its familiar version in ordinary
phase space allows us to directly import formulae that are familiar in the latter context
into the superoperator scenario. For instance, the trace of a superoperator, $\supp{S}$,
takes on the alternative forms:

\ba 
\Tr~\supp{S}  & = \frac{1}{2\pi\hbar}\int \rmd^2 \x~ \dmatel{\opref{\x}}{\supp{S}}{\opref{\x}}
=  \frac{1}{(2\pi\hbar)^2}\int \rmd^2 \x \rmd^2 \bxi~ \supp{S}(\x, \bxi)
\nonumber\\
                &= \frac{1}{2\pi\hbar}\int \rmd^2 \bxi~ \dmatel{\optr{\bxi}}{\supp{S}}{\optr{\bxi}}
= \widetilde{\supp{S}}(\x=0, \bxi=0).               
\ea
Again, the trace of a product of superoperators may be expressed as
\ba
\fl
\Tr~\supp{S_2}~\supp{S_1}
   &= \int \frac{\rmd^2 \x_{-}\rmd^2 \x_{+}}{(2\pi\hbar)^2}~ 
\dmatel{\opref{\x_{-}}}{\supp{S_2}}{\opref{\x_{+}}}~\dmatel{\opref{\x_{+}}}{\supp{S_1}}{\opref{\x_{-}}}
= \int\frac{ \rmd^2 \x \rmd^2 \bxi}{(2\pi\hbar)^2}~ \supp{S_2}(\x, \bxi)\supp{S_1}(\x, \bxi)\nonumber\\
\fl    & =\int\frac{ \rmd^2 \bxi_{-}\rmd^2 \bxi_{+}}{(2\pi\hbar)^2}~ 
\dmatel{\optr{\bxi_{-}}}{\supp{S_2}}{\optr{\bxi_{+}}}~
\dmatel{\optr{\bxi_{+}}}{\supp{S_1}}{\optr{\bxi_{-}}}
= \int \rmd^2 \x \rmd^2 \bxi~ \widetilde{\supp{S_2}}(\x, \bxi)\widetilde{\supp{S_1}}^\ast(\x, \bxi).
\ea
The Wigner-Weyl expressions of partial traces over subsystems can also be immediately generalized
from their versions in single phase space, which are discussed in \cite{LNP}.

%%%%%%%%%%%%%%%%%%%%%%%%%%%%%%%%%%%%%%%%%%%%%%%%%%%%%%%%%%
  \section {Generality of the double Weyl transform}
In the previous section the double Weyl transform was implemented by choosing the conjugate lagrangian planes $Q,P$ (or $\x,\bxi)$
in \eref{xtoQP}. We now show that many other equivalent definitions can be defined using alternative coordinates. Take for example the coordinates $Q_1,P_1$ in \eref{Q1} and the operator basis
\be
\hat Q_\bfa\equiv\dket{Q_\bfa}= \ket{q_{+}} \bra{q_{-}} ,~~~~
\hat P_\balpha\equiv\dket{P_\balpha}= \ket{p_{+}} \bra{-p_{-}},
\ee
$\bfa=(q_+,q_-),\balpha=(p_+,-p_-)$ are conjugate position and momentum variables in double phase space. Starting again from the  {\it double position basis} $\dbra{Q_{\bfa}}\supp{S}\dket{Q_{\bfa'}}$  an alternative {\it double Weyl transform} in this basis is obtained in exact analogy to \eref{dweyl1} as
\be
\supp{S}(\bfa,\balpha)=
\int\rmd^2\bfa'~\dmatel{Q_{\bfa+\half\bfa'}}{\supp{S}}{Q_{\bfa-\half\bfa'}}~\omega^{\balpha\cdot\bfa'}
\label{dweyl2}
\ee
Again a reflection superoperator results as
\be
\supp{R}'(\bfa,\balpha)=\int\rmd^2\bfa'~\dket{Q_{\bfa+\half\bfa'}}\dbra{Q_{\bfa-\half\bfa'}}~\omega^{-\balpha\cdot\bfa'}
\label{refab}
\ee
where now the reflection is in the position basis $\dket{Q_\bfa}$. In this case the Choi conjugation \eref{relinv} between the two position bases  consists of a simple transposition of indices 
\be
\dket{Q_{a,b}}\dbra{Q_{c,d}}=\hat Q_{a,c}\bullet \hat Q_{b,d}
\ee
Applying this transformation to \eref{refab} 
\be
\supp{R}'(\bfa,\balpha)=\int\rmd q'_+\rmd q'_- \ket{q_++\frac{ q'_+}{2}}\bra{q_+-\frac{ q'_+}{2}}\bullet \ket{q_--\frac{ q'_+}{2}} \bra{q_-+\frac{ q'_+}{2}}\omega^{p_-q'_--p_+q'_+}
\ee
we recognize again the appearence of the reflection superoperator $\oprefh{\x_+}\bullet \oprefh{\x_-}$.Thus
\be
\supp{R}'(\bfa,\bfalpha)=\oprefh{\x_+}\bullet \oprefh{\x_-}
\ee
where $(\bfa,\bfalpha)=(q_+,q_-,p_+,-p_-)$ is related to $(\x_+ ,\x_-)=(q_+,p_+,q_-,p_-)$ by a signed permutation matrix ${\cal U}_1$ that satisfies \eref{condition}.  Just as in \eref{centerpm} we then obtain again
\be
\supp{S}'(\bfa ,\bfalpha)={ C}_\supp{S}(\x_+ ,\x_- ).
\label{centerab}
\ee
Thus a rotation of the arguments of the Choi matrix (by a linear transformation ${\cal U}_1$ satisfying \eref{condition}) results in the double Wigner transform of $\supp{S}$ in a {\it different basis}.
Of course, the reversal of the Weyl transform \eref{invref} can also be used to retrieve the
double position or the double momentum representation:
\be
\dmatel{Q_{\bfa +\half \bfa'}}{\supp{S}}{Q_{\bfa -\half \bfa'}}=\frac{1}{(2\pi\hbar)^2}\int \rmd^2 \bfalpha~\supp{S}'(\bfa,\bfalpha)~ \omega^{-\bfa' \cdot \bfalpha}.
\label{invref2}
\ee
Moreover, comparing \eref{centerab} with \eref{centerpm} we also obtain
\be
\supp{S}'(\bfa,\bfalpha)=\supp{S}(\x,\bxi)=C_\supp{S}(\x_+,\x_-).
\ee
In short, we obtain that the single function ${ C}_\supp{S}(\x_+,\x_-)$ supports the Weyl transform of $\supp{S}$ in two different bases just as would be expected from symplectic invariance for the ordinary Weyl transform in single phase space. We note here that the {\it classical} transformation $(\bfa,\bfalpha)\mapsto(\x,-J\bxi)$ preserves the symplectic form in 4D and  therefore belongs to the symplectic group Sp(4). 
An almost identical derivation involving the bases $\dket{Q_2}=\ket{p_+}\bra{q_-},\dket{P_2}=\ket{q_+}\bra{p_-}$ yields 
$\supp{S}''(Q_2,P_2)={ C}_\supp{S}(\x_+,\x_-)$. Clearly an analogous discussion relates the various double chord transforms to the Choi matrix $\widetilde{ C}_\supp{S}(\x_+,\x_-)$ in the translation basis.

Indeed, the immediate generalization
of the above discussion would show that the transit between the various operator representations
corresponding to different choices of double phase space coordinates \eref{classgenfun} among themselves (as well as with the Weyl and the chord representations) is universally obtained from
a single double Weyl representation by a symplectic change of coordinates. To complete the generalization one would need to attempt the explicit construction of metaplectic superoperators in double phase space that would implement the change of basis. We intend to proceed along this path in future publications.
%%%%%%%%%%%%%%%%%%%%%%%%%%%%%%%%%

%%%%%%%%%%%%%%%%%%%%%%%%%%%%%%%%%%%%%%%%%%%%
\section{Quantum evolutions}

Consider the superoperator
\be
\supp{U}=\hat U\bullet \hat U^\dagger,
\ee
which propagates unitarily the density matrix as $\hat U\hat \rho~ \hat U^\dagger$. The matrix elements in the reflection basis are $\dmatel{\opref{\x_+}}{\supp{U}}{\opref{\x_-}}$. This is the kernel that propagates unitarily the Weyl representation of the density matrix $\dbraket{\opref{\x}}{\rho}$
\be
\dbraket{\opref{\x_+}}{\rho'}=\frac{1}{2\pi\hbar}\int \rmd^2\x_-~\dmatel{\opref{\x_+}}{\supp{U}}{\opref{\x_-}}\dbraket{\opref{\x_-}}{\rho}.
\ee
As a particular case of the operators treated in section 3, 
the Choi matrix of $\supp{U}$ is easily computed in a separable form as $ C_\supp{U}(\x_+,\x_-)= U(\x_+)U^\ast(\x_-)$ where $U(\x)=\dbraket{\opref{\x}}{U} $ is the Weyl transform of $\hat U$,
i.e. the {\it Weyl propagator}. The double Weyl transform of $\supp{U}$ is then given by \eref{dweyl2} as $\supp{U}(\x,\bxi)= \Tr\supp{U}\supp{R}_{\x,\bxi}= U(\x+\half\bxi) U^\ast(\x-\half\bxi)$.
The double propagator can then be computed explicitly using \eref{invref}: 
\be
\dmatel{\opref{\x+\half\x_1}}{\supp{U}}{\opref{\x-\half\x_1}}=\frac{1}{2\pi\hbar}
\int \rmd^2 \bxi~U(\x+\half\bxi)U^\ast(\x-\half\bxi)\omega^{\sym{\x_1}{\bxi}}.
\label{invref3}
\ee
At first sight, this centre-centre propagator seems to result from a double Weyl transform 
from the single Weyl propagator, $U(\x)$, of the unitary map $\hat U$ (see \cite{RiosOA, Dittrich}),
but the derivation of \eref{invref} clarifies its true role as the inverse transform from the double Wigner function, that is the Choi representation of the super evolution operator. 
\footnote{One should note that this product of simple Weyl propagators in the integrand was mistakenly identified in \cite{AlmBro11} with the {\it mixed centre-chord propagator}, defined in \cite{AlmBro06}.}
In the semiclassical regime the ordinary Weyl propagators have explicit formulae in terms of generating functions \cite{Ber89, Report}. Further evaluation by stationary phase leads to semiclassical approximations for this propagator. Very similar formulae for the {\it chord-chord propagator} 
can be derived in the translation basis: $\dmatel{\optr{\bxi_{+}}}{\supp{U}}{\optr{\bxi_{-}}}$. 

A more general evolution is generated by the Kraus superoperator \cite{Kraus},
\be
\supp{K}= \sum_j {\hat K}_j \bullet {\hat K_j}^\dagger
\ee
which is again easily accommodated in the present framework. Indeed, linearity 
of the Fourier transforms then specifies the integral kernel for evolving Wigner functions as
\be
\fl \dmatel{\opref{\x+\half\x_1}}{\supp{K}}{\opref{\x-\half\x_1}}=\frac{1}{2\pi\hbar}
\sum_j \int \rmd^2 \bxi~K_j(\x+\half\bxi){K_j}^\ast(\x-\half\bxi)\omega^{\sym{\x_1}{\bxi}}.
\label{invref4}
\ee

An alternative generalization from ordinary unitary evolution, 
which is included in the general formulae of the previous section, is the superoperator
\be
\supp{U}=\hat U_1\bullet {\hat U_2}^\dagger.
\ee
This determines the evolving kernel for the quantum fidelity, or the quantum Loschmidt echo
\cite{Gorin,ZOA11,IFVR}, that is, the overlap of two different evolutions for the same initial state.
Unlike the previous examples, the trace of the evolving operator is not preserved and the identity
operator is not invariant.

\begin{figure}
\begin{center}
\includegraphics[width=6 cm]{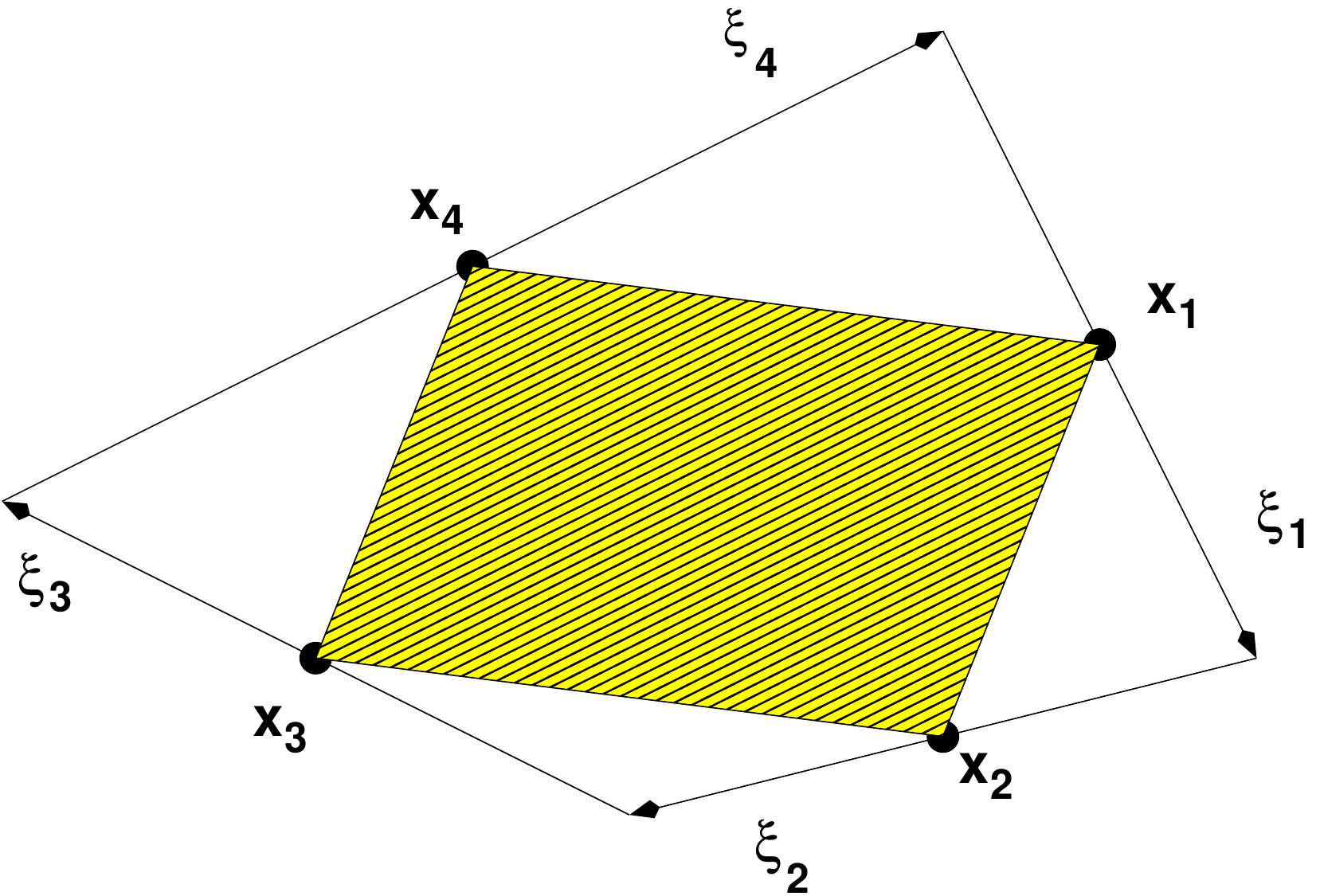}
\caption{The symplectic area of the quadrilateral, $\Delta_4(\bxi_1,\bxi_2,\bxi_3,\bxi_4)=\half(\sym{\bxi_1}{\bxi_2}+\sym{\bxi_3}{\bxi_4})$,
 determines the phase for the action of the superoperator $\optrh{\bxi_4}\bullet\optrh{\bxi_2}^\dagger$
and the superoperator $\oprefh{\x_4}\bullet\oprefh{\x_2}$. It is also identified with the phases for the
trace of the product of four reflections or four translations that are evaluated in the Appendix.}
\label{phases}
 \end{center}
\end{figure}

The super-reflection, $\supp{R}_{\x,\bxi}$, in \eref{suprefx} is precisely of this form 
(within factors of two), choosing $\hat U_1=\oprefh{\x+\half\bxi}$ and
$\hat U_2=\oprefh{\x-\half\bxi}$. The same goes for the super-translation, $\supp{T}_{\x,\bxi}$, 
defined in \eref{suptransx}. As revealed by \eref{actR} and \eref{actT}, 
these superoperators can be viewed as active agents, rather than mere passive Choi bases. 
For instance, the general rules for products of reflections
and translations, which are reviewed in the Appendix, determine the action of a super-translation,
$\optrh{\bxi_4}\bullet\optrh{\bxi_2}^\dagger$ acting on the translation operator $\optrh{\bxi_3}$ 
as just $\omega^{\Delta_4} \optrh{-\bxi_1}$, with $\bxi_1=\bxi_2+\bxi_3+\bxi_4$ as depicted in \Fref{phases}. Likewise, the action of the superoperator $\oprefh{\x_4}\bullet\oprefh{\x_2}$ on the reflection $\oprefh{\x_3}$ is simply $\omega^{\Delta_4}\oprefh{\x_1}$, with $\x_1=\x_2=\x_4-\x_3$.
Note that the four reflection centres $\x_j$ form a parallelogram with half the symplectic area of the circumscribed quadrilateral, $\Delta_4$, formed by the translation chords, $\bxi_j$,
so one obtains the same phase for the composition of three translations or three reflections. Indeed, this is also the same phase as determines the traces
of four translations or four reflections that are discussed in the Appendix.
%%%%%%%%%%%%%%%%%%%%%%%%%%%%%%%%%%%%%%%%%%%%%%%%%%%%%%%%%%%%%%%%%%%%%%%%
%%%%%%%%%%%%%%%%%%%%%%%%%%%%%%%%%%%%%%%%%%%%%%%%%%%%%%%%%%%%%%%%%%%%%%%%%%%

The extent to which \eref{suprefx} and \eref{suptransx} are natural definitions for the reflection and translation superoperators is iluminated by considering the graphs of classical translations and reflections as planes in double phase space. First one notes that each operator, 
$\optrh{\bxi_3}$, corresponds to the translation, $\x_{-} \mapsto \x_{+}=\x_{-}+ \bxi_3$: 
The locus of all pairs $(\x_{-},\x_{+})$ in a given initial plane in double phase space. 
Then the pair of translations, $\x_{+} \mapsto \x'_{+}=\x_{+}+ \bxi_4$ and $\x_{-} \mapsto \x'_{-}=\x_{-}- \bxi_2$, that is, the double phase space translation $X \mapsto X'$, 
takes all double phase space points in the original plane to the new plane $\x_{+}=\x_{-}- \bxi_1$. Thus, the new and the old planes representing translations are themselves related by just the double phase space translation corresponding to $\optrh{\bxi_4}\bullet\optrh{\bxi_2}^\dagger$, that is, $\optrh{\bxi_4}\bullet\optrh{\bxi_2}^\dagger(\optrh{\bxi_3})=\omega^{\Delta_4}\optrh{-\bxi_1}$. 
In the same way, one also finds that the combination of phase space reflections within each of the phase spaces $\x_{\pm}$, corresponding to $\oprefh{\x_4}\bullet\oprefh{\x_2}$, 
transport the double phase space points $X=(\x_{-}= \x_3 - \half\bxi_3, \x_{+}= \x_3 + \half\bxi_3)$ 
in an initial reflection, to a new reflection, $\x_1+ \half\bxi_1 \mapsto \x_1- \half\bxi_1$.
%%%%%%%%%%%%%%%%%%%%%%%%%%%%%%%%%%%%%%%%%%%%%%%%%%%%%%%%%%%%%%%%

\section{Phase space correlations and general pure state conditions}

Consider a general quantum state represented by its density matrix
$\oprho$. Its  phase space translation coresponds to the operator
\be
{\oprho}_{\x} = {\opT}_{\x}~\oprho~{\opT^{\dagger}}_{\x}.
\ee
and the (non-normalised) {\it phase space correlation function} \cite{OVS, LNP} 
is naturaly defined as 
\be
C_{\x} \equiv {\rm tr}~ \oprho~{\oprho}_{\x}=\tr(\oprho{\opT}_{\x}~\oprho~{\opT^{\dagger}}_{\x}).
\label{corr}
\ee
Likewise, from the reflected state ${\oprho}^{\x}= {\opR}_{\x}~\oprho~{\opR}_{\x}$, 
an analogous {\it anticorrelation function} can be defined as
\be
C^{\x} \equiv {\rm tr}~ \oprho~{\oprho}^{\x}=\tr( \oprho~{\opR}_{\x}~\oprho~{\opR}_{\x}).
\label{anticorr}
\ee
To provide a link with our previous results it is then convenient to define 
the {\it density superoperator},
\be
\supp{P}\equiv \oprho \bullet \oprho,
\ee
in terms of which one can identify
\be
C_{\x} = \dmatel{\optr{\x}}{\supp{P}}{\optr{\x}}
~~~{\rm and}~~~ 
C^{\x} = \dmatel{\opref{\x}}{\supp{P}}{\opref{\x}}
\ee
as the diagonal matrix elements of $\supp{P}$ in the translation and reflection bases.
Utilizing the cyclic invariance of the trace in \eref{corr} and \eref{anticorr}, 
we can alternatively rewrite
\be
C_{\x} = \dmatel{\rho}{\optr{\x}\bullet\optr{\x}^\dagger}{\rho}
~~~{\rm and}~~~ 
C^{\x} = \dmatel{\rho}{\opref{\x}\bullet\opref{\x}}{\rho}.
\ee
The Choi-conjugate relations \eref{choitr} \eref{choiref} lead to
\ba
C_{\x}&=&\frac{1}{2\pi\hbar}\int \rmd^2 \bxi_1
\dbraket{\rho}{\optr{\bxi_1}}\dbraket{\optr{\bxi_1}}{\rho}\label{Cx1}\\
C^{\x} &=&\frac{1}{2\pi\hbar}\int \rmd^2 \x_1
~\dbraket{\rho}{\opref{\x+\half\x_1}}\dbraket{\opref{\x-\half\x_1}}{\rho},
\label{Cx2}
\ea
that is, both correlations can then be computed directly from the chord and the (ordinary) Wigner function and chord function \eref{defwigchord}. 

In the case of pure states, $\oprho= |\psi\rangle\langle\psi|$, 
the correlations can be computed directly: 
\be
\fl C_{\x} = |\dbraket{\optr{\x}}{\rho}|^2=(2\pi\hbar)^2|\chi(\x)|^2~~~ {\rm and}~~~
C^{\x} = |\dbraket{\opref{\x}}{\rho}|^2=(2\pi\hbar)^2 W(\x)^2
\ee 
and then comparison with \eref{Cx1},\eref{Cx2} leads to the identities \cite{Chountasis, OVS},
\be
|\chi(\x)|^2 =\frac{1}{2\pi\hbar}~\int \rmd^2\y~ \omega^{\sym{\x}{\y}} |\chi(\y)|^2,
\label{Finv1}
\ee
and
\be
W(\x)^2 = \int \frac{\rmd^2 \y}{2\pi\hbar}~W(\x+\frac{\y}{2})~W(\x - \frac{\y}{2}) .
\ee
Further relationships between products of center and chord functions for pure states 
are obtained from the off-diagonal matrix elements of $\supp{P}$, such as the matrix element
\ba
 \dmatel{\optr{\x-\half\bxi}}{\supp{P}}{\optr{\x+\half\bxi}}&&= 
 \dmatel{\rho}{\optr{\x+\half\bxi}\bullet\optr{\x-\half\bxi}^\dagger}{\rho} \nonumber\\
 &&=\frac{1}{2\pi\hbar}\int \rmd^2 \bxi_1
\dbraket{\rho}{\optr{\bxi_1+\half\bxi}}\dbraket{\optr{\bxi_1-\half\bxi}}{\rho}\omega^{\sym{\bxi_1}{\x}}.
\label{rhorho}
\ea

Now we introduce the remarkable property of pure states that the superoperator $\supp{P}$ is {\it self Choi-conjugate}, i.e. $\oprho \bullet \oprho\equiv\dket{\rho}\dbra{\rho}$, as can be easily checked 
by acting on any operator. It is then possible to pull back the Choi-conjugation relations as unsuspected properties of ordinary pure state Wigner or chord functions in single phase space. Indeed the l.h.s. of \eref{rhorho} is then reduced to 
$\dbraket{\optr{\x-\half\bxi}}{\rho}\dbraket{\rho}{\optr{\x+\half\bxi}}$ so that
\be
\chi(\x+\case{1}{2}\bxi)~ \chi(\x-\case{1}{2}\bxi)^\ast= \frac{1}{2\pi\hbar}\int \rmd^2\bxi_1~
\chi(\bxi_1+\case{1}{2}\bxi)~\chi^\ast(\bxi_1-\case{1}{2}\bxi)~\omega^{\sym{\bxi_1}{\x}}
\label{prop1}
\ee
is identified as a general requirement for chord functions of pure states. The previous case of \eref{Finv1} 
is included as $\bxi=0 $. The off-diagonal elements in the reflection basis yield a similar property for Wigner functions,
\be
W(\x+\case{1}{2}\bxi)~W(\x-\case{1}{2}\bxi)=\frac{1}{2\pi\hbar}\int \rmd^2\x_1~
W(\x+\case{1}{2}\x_1)~W(\x-\case{1}{2}\x_1)~\omega^{\sym{\bxi}{\x_1}},
\label{prop2}
\ee
and a mixed case relating Wigner and chord functions is
\be
W(\x+\case{1}{2}\bxi)~W(\x-\case{1}{2}\bxi)=\frac{1}{2\pi\hbar}\int \rmd^2\bxi_1~
\chi(\bxi+\case{1}{2}\bxi_1)~\chi^\ast(\bxi-\case{1}{2}\bxi_1)~\omega^{\sym{\bxi_1}{\x}}.
\label{prop3}
\ee

These formulae establish families of Fourier identities parametrized continuously by $\x$ or $\bxi$.
They show e.g. that the product of symmetricaly displaced Wigner functions, $W(\x+\case{1}{2}\bxi)~W(\x-\case{1}{2}\bxi)$, is its own Fourier transform for all $\x$ and  $\chi(\x-\case{1}{2}\bxi)^\ast~\chi(\x+\case{1}{2}\bxi)$ is likewise invariant for all values of $\bxi$.
Many  particular cases give rise to interesting properties. Besides the ones already noted for the correlations, we list the following special cases
\ba
\chi(\case{1}{2}\bxi)^2=\frac{1}{2\pi\hbar}\int \rmd^2\bxi_1~\chi(\bxi_1+\case{1}{2}\bxi)~\chi^\ast(\bxi_1-\case{1}{2}\bxi)\\
W(\case{1}{2}\bxi)~W(-\case{1}{2}\bxi)=\frac{1}{2\pi\hbar}\int \rmd^2\x_1~
W(\case{1}{2}\x_1)~W(-\case{1}{2}\x_1)~\omega^{\sym{\bxi}{\x_1}}\\
W(\case{1}{2}\bxi)~W(-\case{1}{2}\bxi)=\frac{1}{2\pi\hbar}\int \rmd^2\bxi_1~
\chi(\bxi+\case{1}{2}\xi_1)~\chi(\bxi-\case{1}{2}\bxi_1)\\
W(\x)^2=\frac{1}{2\pi\hbar}\int \rmd^2\bxi_1~
\chi(\case{1}{2}\bxi_1)~\chi^\ast(-\case{1}{2}\bxi_1)~\omega^{\sym{\bxi_1}{\x}}.
\ea
The last one, taking into account that $\chi(\x)=\chi^\ast(-\x)$ and Parseval's relation, yields  the further integral
\be
\int \rmd^2\x~ W(\x)^4=\int \rmd^2\x~ |\chi(\case{1}{2}\x)|^4.
\ee

Furthermore, for $\x=\bxi=0$, one obtains:
\be
W(0)^2=\frac{1}{2\pi\hbar}\int \rmd^2\x~ W(\case{1}{2}\x)W(-\case{1}{2}\x)=\frac{1}{2\pi\hbar}\int \rmd^2\bxi_1~\chi(\case{1}{2}\bxi_1)~\chi^\ast(-\case{1}{2}\bxi_1).
\ee

The way the general formulae \eref{prop1},\eref{prop2},\eref{prop3} were derived implies that they are necessary conditions for pure state distributions. By setting $\x=\case{1}{2}\bxi$, it is easily shown that \eref{prop1} is equivalent to the pure state condition $\hat \rho=\hat \rho^2$ in the chord representation, and so it is also sufficient. The case of the
Wigner function is not so transparent, but the integral product rule for the Weyl representation \cite{Report} for $\hat \rho^2$,
\ba
\rho^2(\x)&&= 4 \int \rmd^2\x_1\rmd^2\x_2~W(\x_1)~W(\x_2)~\omega^{2\sym{(\x_1-\x)}{(\x_2-\x)}}\nonumber\\
&&= 4 \int \rmd^2\bar{\x}~ \int\rmd^2\bxi ~W(\bar{\x}+\case{1}{2}\bxi)~W(\bar{\x}-\case{1}{2}\bxi)~\omega^{2\sym{(\x-\bar{\x})}{\bxi}},
\ea 
is imediately simplified by \eref{prop2}
so that
\be
\rho^2(\x)= 8\pi\hbar~ W(\x)\int \rmd^2\bar{\x}~ W(2\bar{\x}-\x)=2\pi\hbar~ W(\x) = \rho(\x).
\ee

\subsection{Airy functions, an example}

There are some notable cases where the Wigner function of pure states can be described 
in terms of standard special functions found in eg \cite{Abramowitz}. Then the identity,
\begin{eqnarray} 
W(\x+\frac{\y}{2})~W(\x-\frac{\y}{2}) 
=\int \frac{d\x'}{(2\pi\hbar)}~ W(\x+\frac{\x'}{2})~W(\x-\frac{\x'}{2})~\omega^{<\x', \y>}, 
\end{eqnarray}
implies a possibly unsuspected Fourier identity
for a symmetrized product of analytic functions. Such is the case for the eigenfunctions of the harmonic oscillator - given in terms of Laguerre polynomials \cite{Groenewold} - or the unnormalized eigenfunctions of the hyperbolic hamiltonian $H(\x)=p~q$, calculated in \cite{Voros} in terms of Laguerre functions of complex index. We develop here the important example of the linear potential $V(q)= q$.
If $m=1/2$ and $\hbar=1$, so that the Hamiltonian is simply $H(\x) = p^2 + q$, 
the zero energy eigenfunction is proportional to the Airy function, 
\be
{\rm Ai}(q) \equiv \frac{1}{2\pi} \int_{-\infty}^{\infty} dp~ \exp\left[i\left(\frac{p^3}{3}+pq\right)\right],
\ee 
which is not normalizable and hence has no Fourier transform. 
Evidently, the corresponding momentum representation of this state is just
\be
\langle p|\psi\rangle = \frac{1}{\sqrt{2\pi}} \exp\left[i~\frac{p^3}{3}\right],
\ee
so that the corresponding Wigner function is just \cite{BalazsJen}
\begin{eqnarray}
W(\x)&& = \frac{1}{2\pi} \int dp'~ \langle p+\frac{p'}{2}|\psi\rangle 
\langle \psi|p-\frac{p'}{2}\rangle \exp[ip'q] \nonumber \\
&& = \frac{1}{(2\pi)^2} \int dp'~ \exp\left[i\left(\frac{{p'}^3}{12}+(p^2+q)p'\right)\right] \nonumber\\
&& = \sqrt{\frac{2^{1/3}}{\pi}}~ {\rm Ai}\left( 2^{2/3} H(\x) \right),
\end{eqnarray}
which is also not square-integrable. 

Nonetheless, the product $W(\x+\frac{\x'}{2})~W(\x-\frac{\x'}{2})$
decays exponentially in the $\x'$-phase plane outside the region limited by the pair of reflected parabolae, $H( \x\pm\frac{\x'}{2})= 0$, so that it is square integrable.
Thus the direct verification of the Fourier identity \eref{prop2} in the case of the Airy function 
proceeds from the integral representation:
\begin{eqnarray} 
  &&W\left(\x_1 + \frac{\x_2}{2}\right)W\left(\x_1 - \frac{\x_2}{2}\right) = \nonumber \\ 
  &&
\frac{1}{(2\pi)^4} \int dp'dp'' 
\exp\left[\rmi\left(\frac{{p'}^3}{12}+\left((p_1+\frac{p_2}{2})^2+q_1+\frac{q_2}{2}\right)p'\right)\right]
\nonumber \\ && ~~~~~~~~~~~~~~~
\exp\left[-\rmi\left(\frac{{p''}^3}{12}-\left((p_1-\frac{p_2}{2})^2+q_1-\frac{q_2}{2}\right)p''\right)\right].
\end{eqnarray}
Then the transformation $p'= a+\frac{b}{2}, p''= a-\frac{b}{2}$ simplifies this into
\begin{eqnarray} 
&&W\left(\x_1 + \frac{\x_2}{2}\right)W\left(\x_1 - \frac{\x_2}{2}\right) =  \nonumber \\
&&\frac{1}{(2\pi)^4} \int da~ db~
\exp\left[i\left(\frac{{b}^3}{48}+ \frac{a^2 b}{4} + (q_2 + 2p_1 p_2) a + ({p_1}^2+\frac{{p_2}^2}{4}+q_1)b\right)\right].
\end{eqnarray}
%%%%%%%%%%%%%%%%%%%%%%%%%%%%%%%%%%%%%%%%
\begin{figure} 
\begin{center}
\includegraphics[width=6 cm]{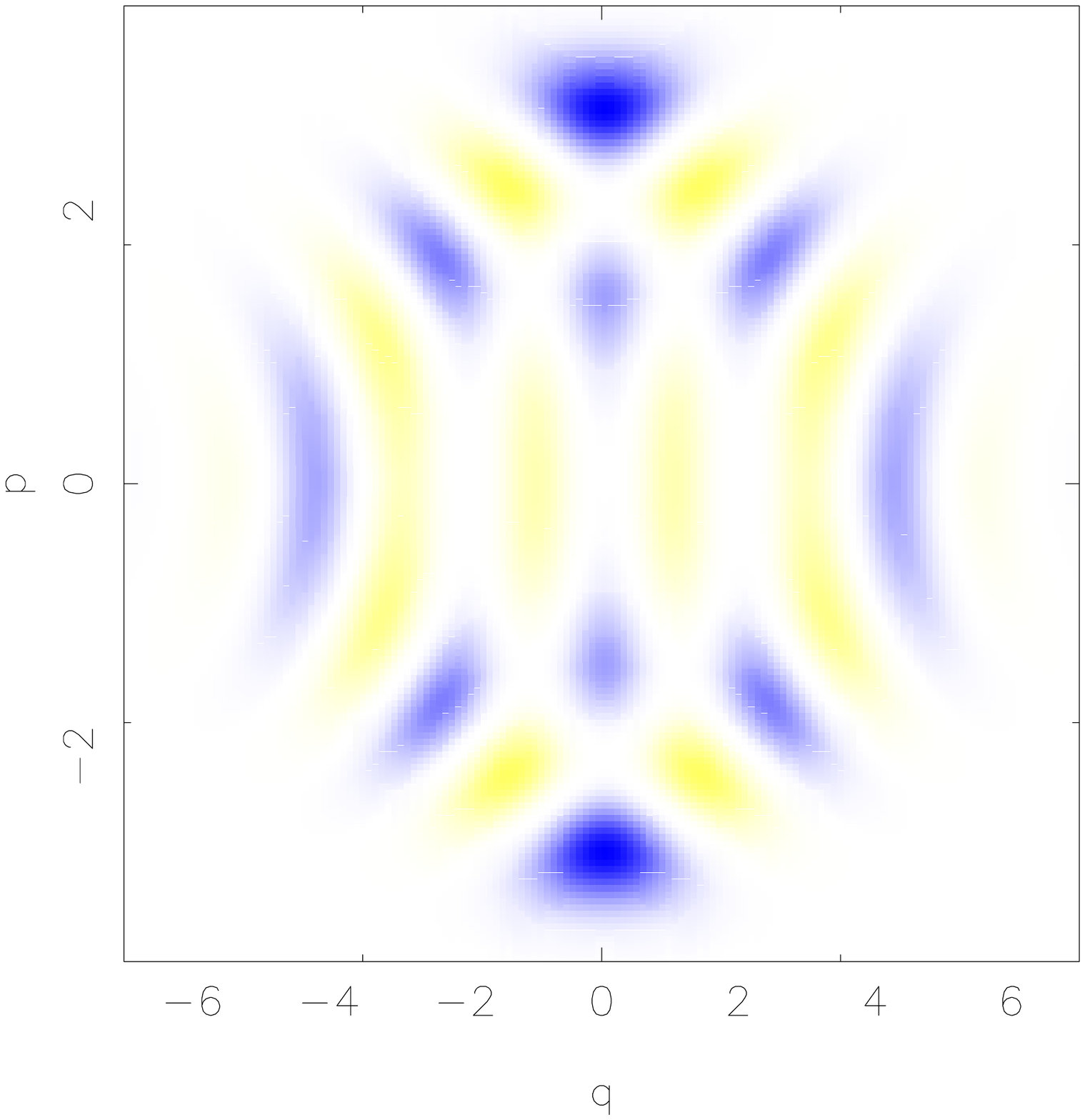}
\caption{The product of two Airy functions as in \eref{airyproduct}for $\x_3$ for $\x_1=(-3,0)$. In the color density plot blue is positive and yellow negative}.
\label{airy}
\end{center}
 \end{figure}
%%%%%%%%%%%%%%%%%%%%%%%%%%%%%%%% 
Thus, the Fourier transform with respect to $\x_2$ becomes
\begin{eqnarray} 
&&\frac{1}{2\pi} \int dx_2~ \e^{i <\x_2, \x_3>}~ 
W\left(\x_1 + \frac{\x_2}{2}\right)W\left(\x_1 - \frac{\x_2}{2}\right) = 
\frac{1}{(2\pi)^5} \int da~ db~dq_2 dp_2 ~ \nonumber \\
&&\exp\left[\rmi\left(\frac{{b}^3}{48}+ \frac{a^2 b}{4} + (q_2 + 2p_1 p_2) a + ({p_1}^2+\frac{{p_2}^2}{4}+q_1)b + p_2 q_3 - q_2 p_3\right)\right] \nonumber \\
&& =\frac{1}{(2\pi)^4} \int db~ dp_2 ~ 
\exp\left[\rmi\left(\frac{{b}^3}{48}+ \frac{{p_3}^2 b}{4} + 2p_1 p_2 p_3 + ({p_1}^2+\frac{{p_2}^2}{4}+q_1)b + p_2 q_3\right)\right] \nonumber \\
&& = W\left(\x_1 + \frac{\x_3}{2}\right)W\left(\x_1 - \frac{\x_3}{2}\right),
\end{eqnarray}
which implies the Fourier invariance for a product of symmetrized Airy functions,
\begin{eqnarray}
&&\frac{1}{2\pi} \int dx_2~ \e^{i <\x_2, \x_3>} 
{\rm Ai}\left( 2^{2/3} H(\x_1 + \frac{\x_2}{2})\right)~
{\rm Ai}\left( 2^{2/3} H(\x_1 -\frac{\x_2}{2})\right) \nonumber \\
&& = {\rm Ai}\left( 2^{2/3} H(\x_1 + \frac{\x_3}{2})\right)~
{\rm Ai}\left( 2^{2/3} H(\x_1 -\frac{\x_3}{2})\right),
\label{airyproduct}
\end{eqnarray}
not encountered even in \cite{ValSo}, a book dedicated specifically to Airy functions or \cite{AbrRaz}. In \Fref{airy} we show this product in the plane $\x_3$ for $\x_1=(-3,0)$. Notice the central symmetry implied by the Fourier invariance.

An important feature of this example is that the semiclassical {\it transitional approximation}
of pure state Wigner functions for general WKB-quantized states near the closed energy eigencurve
was shown by Berry \cite{Berry77} to be just the Airy function over the approximating parabola. 
Thus one finds that both the exact Wigner function and its transitional approximation satisfy the new
Fourier invariance, even though this is not the case of other semiclassical approximations 
that are more refined in other respects.

%%%%%%%%%%%%%%%%%%%%%%%%%%%%%%%%%%%%%%%%%%%%%%%%%%%%%%%%%

\section{Conclusions and outlook}

Two possible representations of a superoperator can be naturally derived from the same operator basis. They are unitarily related, and we have referred to them as Choi-conjugate. We have developed here the general relationship between them - when the basis is orthogonal - and we have studied in particular the form of this relationship when the unitary bases of translations and reflections are used.
It turns out that the representation in terms of the Choi or dynamical matrix $C_{\supp{S}}$ can be interpreted as a double Weyl or Wigner transform of the matrix elements of the superoperator. This is because the representation in terms of the Choi matrix is actually an expansion in terms of translation and reflection {\it superoperators}, in strict analogy to the expansion of an ordinary operator in terms of translations and reflections, yielding its Weyl or Wigner transform. The definition of 
$\supp{T}_{\x,\bxi}$ and $\supp{R}_{\x,\bxi}$ opens up the possibility for a full study of the affine geometry of these superoperators  in double phase space, including the definition of unitary superoperators that implement symplectic transformations belonging to $Sp(4)$. We intend to pursue this analysis in the future.

Our treatment here has been for the simplest case of a phase space with no boundaries and with one degree of freedom. The extension to $D$ degrees is immediate and needs no further comment. The adaptation of our techniques to a phase space with boundaries needs more care.
The case of torus topology - periodic boundary conditions both in position and momentum -is the closest to the present approach and leads to a finite dimensional Hilbert space of integer dimension $d={\rm area}/2\pi\hbar$ \cite{Leonhardt},\cite {Wootters},\cite{Rivas},\cite{Miquel} and is of great current interest in quantum information theory. Translation and reflection operators can still be defined and provide a basis for a similar treatment as the one developed here. In this context, the display in double phase space of the properties of superoperators can provide new insights into their actions, just as the celebrated Wigner and Weyl representation displayed properties of quantum states in single phase space. The action of gaussian noise channels in the chord representation \cite{Aolita} is a first step in that direction.

As application of these methods we have found some previously unknown identities relating products of Wigner and Weyl distributions for pure states. These identities generalize the pure state conditions and in some cases produce new relationships for the special functions of analysis. 

\ack
We thank Raul Vallejos for a careful reading and many comments. Financial support from National Institute for Science and Technology--Quantum Information, FAPERJ and CNPq is gratefully acknowledged.

%%%%%%%%%%%%%%%%%%%%%%%%%%%%%%%%%%%%%%%%%%%%%%%
\appendix
\section{}
We review the well known definitions and properties of reflection and translation operators. They constitute the foundation for the Weyl representation of quantum mechanical operators as phase space c-number functions (the Wigner quasiprobability distribution in the case density matrices).
 We start with the usual $\hat q, \hat p$ operators that we subsume in a phase space operator $\hat {\bf x}=(\hat q,\hat p)$ and a phase space label $\mbfx=(q,p)\in {\mathbb R}^2$ The corresponding position and momentum bases are denoted $\ket{q},\ket{p}$.
The symplectic product is defined as
\be
\sym{\mbfx}{\mbfx'}= (q,p)\matr{0}{-1}{1}{0}\vectr{q'}{p'}= pq'-qp'.
\label{symplectic}
\ee
For notational simplicity we also introduce the quantities 
\be
\tau=\rme^{\rmi/(2\hbar)}~~~~~~, ~~~~ \omega=\rme^{\rmi/\hbar}.
\ee
Reflections and translation operators are defined as 
\ba
\oprefh{\x}\equiv\hat R_{q,p} &=&\int \rmd q' ~\ket{q+\frac{q'}{2}}\bra{q-\frac{q'}{2}}~\omega^{pq'} \\
\optrh{\bxi}\equiv\hat T_{\xi_q,\xi_p} &=&\int \rmd p' ~\ket{p'+\frac{\xi_p}{2}}\bra{p'-\frac{\xi_p}{2}}~\omega^{-\xi_q p'},
\ea
where the first is in the position and the latter in the momentum basis. They have the properties
\be
\optrh{\bxi}^\dagger=\optrh{-\bxi},~~~~~~\oprefh{\x}^\dagger=\oprefh{\x},~~~~~~(\half\oprefh{\x})^2=1.
\ee
Their action on the position and momentum basis justifies their names
\be
\oprefh{\x}\ket{q_0}=2~\ket{2q-q_0}~\omega^{2(q-q_0)p}  ~~~~~~~~~~~~\opref{\x}\ket{p_0}=2~\ket{2p-p_0}~\omega^{-2q(p-p_0)}
\label{actrefl}
\ee 
and
\be
\optrh{\bxi}\ket{p_0}=\ket{p_0+\xi_p}~\omega^{-\xi_q(p_0+\half \xi_p)} ~~~~~~~~~~
\optrh{\bxi}\ket{q_0}=\ket{q_0+\xi_q}~\omega^{\xi_p(q_0+\half\xi_q)}.
\label{acttrans}
\ee
Moreover they form a group which is the representation of the affine group of reflections and translations, with the following composition laws
\ba
\optrh{\bxi_1}\optrh{\bxi_2}&=&\tau^{\sym{\bxi_1}{\bxi_2}}\optrh{\bxi_1+\bxi_2}~~~~~~~~~~~~~~~~
\oprefh{\x_1}\oprefh{\x_2}=4\omega^{2\sym{\x_1}{\x_2}}\optrh{2(\x_2-\x_1)}\\
\oprefh{\x}\optrh{\bxi}&=&\omega^{-\sym{\x}{\bxi}}\oprefh{\x-\bxi/2} ~~~~~~~~~~~~~~~~~\optrh{\bxi}\oprefh{\x}=\omega^{\sym{\bxi}{\x}}\oprefh{\x+\bxi/2}.
\label{group}
\ea
They conform a pair of complementary orthonormal bases with the properties
\be
\tr \optrh{\bxi}^\dagger\optrh{\bbeta}=2\pi\hbar \delta(\bxi-\bbeta), ~~~~
\tr\oprefh{\x}\oprefh{\y}=2\pi\hbar \delta(\x-\y),  ~~~~
\tr\oprefh{\x}\optrh{\bxi}=\omega^{-\sym{\x}{\bxi}}.
\ee
Switching to the double Dirac notation we rewrite the above as
\be
\dbraket{\optrh{\bxi}}{\optrh{\bbeta}}=2\pi\hbar \delta(\bxi-\bbeta),  ~~~~~
\dbraket{\oprefh{\x}}{\oprefh{\y}}=2\pi\hbar \delta(\x-\y),   ~~~~~
\dbraket{\oprefh{\x}}{\optrh{\bxi}}=\omega^{-\sym{\x}{\bxi}}.
\ee
The labels $\x$ and $\xi$ are related to the conjugate variables $Q,P$ of \eref{QPtox} as
$\x=Q$ and $\xi=JP$. Thus we can think of reflection and translation operators as 
alternative position and momentum bases in double phases space.
Using these properties  we compute the quadruple traces needed in the main text: 
\ba
\tr(\optrh{\bxi_1}\optrh{\bxi_2}\optrh{\bxi_3}\optrh{\bxi_4})=2\pi\hbar\delta(\bxi_1+\bxi_2+\bxi_3+\bxi_4)
\tau^{\sym{\bxi_1}{\bxi_2}+\sym{\bxi_3}{\bxi_4}}\\
 \tr(\oprefh{\x_1}\oprefh{\x_2}\oprefh{\x_3}\oprefh{\x_4})=2\pi\hbar\delta(\frac{\x_1+\x_3}{2}-\frac{\x_2+\x_4}{2})
\omega^{2\sym{\x_1}{\x_2}+2\sym{\x_3}{\x_4}}.
\label{traces}
\ea
We should remark at this point that traces of unitary operators are related semiclassically to classical periodic orbits and their actions \cite{Gutzwiller, livro}, even when the classical and quantum evolutions
are broken up into several steps \cite{OABro15}. In the case of four operators $\optrh{\mbfx}$, such a trajectory is composed of four segments, each giving a phase space translation and forming a closed quadrilateral. The action of this trajectory is the symplectic area of the quadrilateral $\Delta_4(\bxi_1,\bxi_2,\bxi_3,\bxi_4)=\half(\sym{\bxi_1}{\bxi_2}+\sym{\bxi_3}{\bxi_4})$ (in units of $\hbar$). In the case of reflections the trajectory connects the centers of the segments of this quadrilateral, which is a parallelogram $\mbfx_1- \mbfx_2 +\mbfx_3 -\mbfx_4=0$. Both cases
are illustrated in \Fref{phases}.
%

%%%%%%%%%%%%%%%%%%%%%%%%%%%%%%%%%%%%%%%%%%%%%%%%%%%%%%%%
\end {document}